\documentclass[12pt,preprint]{aastex}

\def\gta{ \lower .75ex \hbox{$\sim$} \llap{\raise .27ex \hbox{$>$}} }
\def\lta{ \lower .75ex\hbox{$\sim$} \llap{\raise .27ex \hbox{$<$}} }
\begin{document} 
 
\title{The collimation--corrected GRB energies correlate with the peak 
energy of their $\nu F_{\nu}$ spectrum} 
 
\author{Giancarlo Ghirlanda\altaffilmark{1},  
Gabriele Ghisellini\altaffilmark{1} and Davide Lazzati\altaffilmark{2}} 
\affil{1 INAF - Osservatorio  Astronomico di  Brera, via  Bianchi 46, 
23807 Merate,  Italy}  
\affil{2 Institute of  Astronomy, Madingley Road CB3 0HA, Cambrige UK} 
 
 
\setcounter{footnote}{0} 
 
\begin{abstract} 
We  consider all  bursts  with  known redshift  and  $\nu F_\nu$  peak 
energy,  $E^{obs}_{peak}$. 
For  a good fraction  of them an  estimate of the  jet opening 
angle is available from the  achromatic break of their afterglow light 
curve.   This  allows  the  derivation of  the  collimation--corrected 
energy of the  bursts, $E_\gamma$.  The distribution of  the values of 
$E_\gamma$ is more spread  with respect to previous findings, covering 
about  two  orders  of   magnitude.   We  find  a  surprisingly  tight 
correlation  between  $E_\gamma$  and  the  source  frame  $E_{peak}$: 
$E^{obs}_{peak}(1+z)  \propto E_\gamma^{0.7}$.   This  correlation can 
shed light on  the still uncertain radiation processes  for the prompt 
GRB emission.   More importantly, if  the small scatter of  this newly 
found correlation  will be confirmed  by forthcoming data, it  will be 
possible to use it for cosmological purposes. 
\end{abstract} 
 
\keywords{Gamma Rays: bursts --- Radiative processes:  non-thermal}

\section{Introduction} 
 
The possibility  that GRB fireballs are collimated  was first proposed 
for  GRB~970508 (Waxman  et al.   1998) and  subsequently  invoked for 
GRB~990123  as a  possible explanation  for its  extraordinarily large 
isotropic energy  (Fruchter et al. 1999).   The observational evidence 
supporting  this scenario  is the  achromatic break  of  the afterglow 
light curve  which declines  more steeply than  in the  spherical case 
(e.g.   Rhoads  1997,  Sari,   Piran  \&  Halpern  1999).   Under  the 
simplifying assumption  of a  constant circum-burst density  medium of 
number density  $n$, a fireball  emitting a fraction  $\eta_\gamma$ of 
its  kinetic energy  in the  prompt $\gamma$--ray  phase would  show a 
break  in its  afterglow  light  curve when  its  bulk Lorentz  factor 
$\Gamma$ becomes of the order of $\Gamma\simeq 1/\theta$ with $\theta$ 
given by  (Sari et al.   1999) {\footnote{The notation  $Q=10^xQ_x$ is 
adopted, with cgs units.}}: 
\begin{equation} 
\theta=0.161 \left({ t_{jet,d} \over 1+z}\right)^{3/8} 
\left({n \, \eta_{\gamma}\over E_{\gamma,iso,52}}\right)^{1/8} 
\label{theta} 
\end{equation} 
where $z$ is the redshift, $t_{jet,d}$  is the break time in days, and 
$E_{\gamma,iso}$ is  the energy in  $\gamma$--rays calculated assuming 
that the emission is  isotropic.  The collimation--corrected energy is 
$E_\gamma=(1-\cos\theta)E_{\gamma, iso}$. 
 
Frail  et al.  (2001  -- F01  thereafter) considering  a sample  of 15 
bursts with redshift and estimate of $\theta$ (including 5 lower/upper 
limits)  found the  remarkable result  of a  clustering  of $E_\gamma$ 
around $5\times  10^{50}$ erg.  This was  also confirmed independently 
by Panaitescu \&  Kumar (2001).  More recently, Bloom  et al. (2003 -- 
B03  hereafter)  found that  the  distribution  of  $E_\gamma$ for  a 
larger sample of 24 bursts (including 8 lower/upper limits) clusters 
around  $1.3\times 10^{51}$  erg,  emphasizing at  the  same time  the 
presence  of  several outliers  with  a  sizeably smaller  energetics. 
These  results suggest  that  GRBs are  characterized  by a  universal 
energy reservoir, despite the very large range of isotropic equivalent 
energetics. 
 
A  correlation between  $E_{\gamma, iso}$  and the  source  frame peak 
energy  $E_{peak}$ was  discovered by  Amati et  al. (2002)  (see also 
Lloyd-Ronning et al.  2000, Lamb et al.  2003; Sakamoto et al.  2004). 
An analogous  correlation holds between the  peak luminosity $L_{peak, 
iso}$ and $E_{peak}$ (Schaefer  2003; Yonetoku et al. 2003).  Possible 
interpretations  of  these  correlations  (Schaefer 2003;  Eichler  \& 
Levinson 2004, Liang  et al.  2004) are now  under intense discussion. 
It seems  therefore that the local  energy content of the  jet plays a 
crucial role in determining  the typical photon energy $E_{peak}$.  In 
this paper  we show that  the tightest correlation is  instead between 
$E_{peak}$ and $E_\gamma$. 
 
\section{Sample selection} 
 
We  considered all  the  40  GRBs with  measured redshift 
up to June 2004
\footnote{ A continuously updated collection of GRBs with links to the 
relative      GCN     communications      can     be      found     at 
http://www.mpe.mpg.de/\~jcg/grbgen.html}. 
 
The  prompt  emission   spectrum  of  GRBs  can  be   described  by  a 
phenomenological spectral model (i.e.   the Band function, Band et al. 
1993) composed by low and  high energy powerlaws, with photon spectral 
indices $\alpha$  and $\beta$  respectively, smoothly connected  by an 
exponential cutoff  with a characteristic energy  $E_{break}$.  In the 
$\nu  F_{\nu}$  representation  this  model  predicts  a  peak  energy 
$E_{peak}=(\alpha+2)E_{break}$.   For  all  the bursts  with  measured 
redshift  we searched in  the literature  any information  about their 
prompt emission  spectrum.  All bursts with  firm redshift measurement 
and published $E_{peak}$ were included in our final sample reported in 
Tab.~1.   This sample  contains  29 GRBs  detected  by different 
instruments  (Col.~1)  and distributed  in  redshift (Col.~2)  between 
$z$=0.0085 (GRB~980425, the second  nearest is the X--ray Flash 020903 
with  $z$=0.25,  Soderberg  et  al.  2004)  and  $z$=4.5  (GRB~000131, 
Andersen et al.  2000). Col.~3  to Col.~8 report the parameters of the 
spectrum integrated over the burst duration (Col.~6).  The average low 
(Col.~3) and high  (Col.~4) energy spectral indices of  our sample are 
$\langle\alpha\rangle=-1.05\pm                0.44$                and 
$\langle\beta\rangle=-2.28\pm0.25$,  which  are  consistent  with  the 
typical values found from the  spectral analysis of bright long bursts 
(see, e.g. Preece et al. 2000).  All the GRBs for which a redshift has 
been   measured   belong   to    the   population   of   long   bursts 
(e.g. Kouveliotou et al. 1993)  and the average duration of the sample 
reported in  Tab.~1 is 88  sec.  The fluences (although  not uniformly 
integrated  over the  same  energy range)  are  representative of  the 
typical GRB values.  For the GRB detected by $Beppo$SAX and studied 
by Amati et  al. (2002) the spectral parameters  are derived combining 
the Wide  Field Camera  (WFC) and the  Gamma Ray Burst  Monitor (GRBM) 
instruments  on-board  $Beppo$SAX, while  the  fluence  is  given by  the 
analysis of the  GRBM data only.  There is  a small difference between 
the $\gamma$--ray fluences  derived fitting the GRBM data  only or the 
GRBM+WFC  data.   Therefore,  using   the  spectral  indices  and  the 
normalization obtainable  from the fluences  given in Tab. 1  does not 
lead to the correct value  of $E_{\gamma,iso}$ reported in Tab. 2 (for 
$Beppo$SAX  bursts only).  The  latter is  taken directly  from Amati  et al. 
(2002), i.e.  from the combined fit.  Note also that for bright bursts 
the GRBM instrument is sensitive over a larger energy range (extending 
towards higher energy) than the nominal 40--700 keV interval.
 
Eight  GRBs (Tab.~3)  were excluded  from this  sample  either because 
their redshift was uncertain (2 GRBs) or because their peak energy was 
not found in the literature (6 GRBs).  They are discussed in Sect. 4.1 
and their consistency  with the conclusions drawn for  the main sample 
is checked under reasonable assumptions for the lacking parameters. 
 
From the original  sample of  40 GRBs with  known redshift only 3 
cases were  completely excluded from the  present analysis: GRB~020124 
(Ricker G.  et  al.  2002), whose spectral properties  are reported in 
Barraud et al.  (2003), presents a peak energy of $\sim$ 1 MeV with an 
uncertainty  of an  order of  magnitude; GRB~030323  (Graziani  et al. 
2003) for  which  no  peak  energy  or fluence  was  reported  in  the 
literature and GRB~031203 (Gotz et al. 2003) without a (yet) published 
prompt phase spectrum. 
 
Considering all the bursts presented  in this work (Tab.~1 and Tab.~3) 
our sample contains 37 GRBs.

\section{Isotropic energy and collimation correction} 
 
The source frame isotropic equivalent energy of a burst can be derived 
from its frequency and time integrated flux.  If the jet opening angle 
is  known then  we  can calculate  the  collimation corrected  energy. 
These quantities  for the  sample of selected  bursts are  reported in 
Tab.~2. 
 
Similarly to  Bloom et al.  (2001,  2003) and Amati et  al.  (2002) we 
derived  for   every  burst  reported  in  Tab.~1   the  source  frame 
``bolometric'' isotropic energy  $E_{\gamma,iso}$ integrating the best 
fit time--integrated  model spectrum N(E)  [phot cm$^{-2}$ keV$^{-1}$] 
over  the energy range  1 keV--10  MeV.  The  integration over  such a 
large band also required  the proper correction for the band--redshift 
effect so that: 
\begin{equation} 
E_{\gamma,iso}={4\pi\,D_{l}^{2}\over(1+z)}\, 
\int_{1/1+z}^{10^{4}/1+z}E\,N(E)\,dE\ 
\ \rm erg 
\end{equation} 
where $E$ is in keV and  $D_{l}$ is the source luminosity distance (we 
adopt  $\Omega_{m}$=0.3, $\Omega_{\Lambda}$=0.7, H$_0$=70  km s$^{-1}$ 
Mpc$^{-1}$).  Note that  Amati et al.   2002 assume for  the $Beppo$SAX 
bursts  a value  of H$_0=65$  km s$^{-1}$  Mpc$^{-1}$  in deriving 
$E_{\gamma, iso}$.   For these  bursts we corrected  their $E_{\gamma, iso}$
(col.~7 in  Tab.~2)  for our  different  $H_0$.
 
For 24/29 GRBs reported in  Tab.~1 a jet break time is known and 
the collimation angle  $\theta$ can be derived (Eq.~1).   For these we 
used  the isotropic  energy  estimated above  with  a constant  energy 
conversion efficiency $\eta_\gamma$=20\% (see  also F01).  Only in few 
cases the circum-burst density  was measured (from broad band modeling 
of the afterglow  emission, e.g.  Panaitescu \& Kumar  2001, 2002) and 
used  in deriving  $\theta$.  For  those  bursts with  unknown $n$  we 
assumed the median value $n \simeq 3$ cm$^{-3}$ of the distribution of 
the measured  densities  which extends roughly  between 1  and 10 
$cm^{-3}$ (Frail et al. 2000a, Yost et al.  2002, Panaitescu \& Kumar 
2002,  Frail et  al.  2003, Harrison  et  al.  2001,  Schaefer et  al. 
2003).   This  density  was   also  assumed  for  GRB~990123  although 
Panaitescu \& Kumar (2001), fitting simultaneously the afterglow light 
curves in  different bands,  found for this  burst and  for GRB~980703 
$n\simeq  10^{-3}$  cm$^{-3}$.    Studying  the  radio  properties  of 
GRB~980703,  Frail et  al.   (2003) found  $n\sim  30$ cm$^{-3}$,  and 
discussed  the possible  reasons for  the discrepancy  with  the value 
quoted above (see  also Berger et al.  2004a),  which could be applied 
also to GRB~990123 (see also Nakar \& Piran 2004).  Similar to F01 and 
B03  we  have  then  treated   GRB~990123  as  a  burst  with  unknown 
circum-bursts density,  and therefore used $n=3$  cm$^{-3}$ (note that 
F01 used $n=0.1$ cm$^{-3}$, while B03 used $n=10$ cm$^{-3}$). 
 
The jet break  time (collected in most cases from  B03) is reported in 
Tab.~2 (Col.~1) for all the bursts  (16) for which a direct measure of 
this  parameter was  possible  from  the broad  band  modeling of  the 
afterglow light  curve.  For  7 GRBs only upper/lower  limits on 
$t_{jet}$  were  estimated  from  the available  data.   These  limits 
determine the upper/lower  limits on the collimation--corrected energy 
reported in Tab.~1 (Col.~7).  On the other hand there are a handful of 
GRBs whose  spectrum is best fit  by a Band function  with high energy 
spectral index $\beta>$-2.  This  indicates that $E_{peak}$ is outside 
(above) the fitting  energy range.  These cases are  reported as lower 
limits in  energy (Tab.~2, Col.~8). Only  for XRF 030723  the limit on 
the peak energy and on the isotropic energy is due to the limit on its 
redshift  $z<$2.1 (Fynbo et al. 2004). 
 
The jet  break time, the derived  jet opening angle,  the source frame 
isotropic   energy  $E_{\gamma,iso}$,   collimation--corrected  energy 
$E_\gamma$ and peak  energy $E_{peak}$ are reported in  Tab.~2 for the 
sample of bursts  in Tab.~1.  Tab.~4 reports these  quantities for the 
few GRBs  of Tab.~3 with uncertain  parameters.  We note  that, due to 
the larger  energy range for  the calculation of the  isotropic energy 
and to the slightly different circum-burst assumed density, the values 
of  $\theta$ and  $E_\gamma$  reported in  Tab.~2  are different  with 
respect to those reported by F01 and B03 for the common GRBs.  
 
The  spectral parameters  of several  bursts reported  in  Tab.~1 (and 
Tab.~2)  were  published  without  the  relative  errors.   For  these 
parameters we assumed the average errors (in square parenthesis) 
obtained from those bursts  with published parameters' errors (in 
round  parenthesis).  Furthermore,  we do  not know  the correlations 
among the  parameters entering in  the calculation of the  jet opening 
angle and of  the collimation corrected energy.  For  this reason (see 
also B03)  the simplest approach is  to propagate the  errors in their 
calculations assuming that these  parameters are uncorrelated.  In all 
the  tables  the errors  obtained  with this  method  are 
reported in  square parenthesis and the  resulting average uncertainty 
on  the   three  foundamental  quantities   $E_{peak}$,  $\theta$  and 
$E_{\gamma}$ are 20\%, 32\% and 20\%, respectively.

\section{Results} 
 
In  Fig.~1 we  report ({\it  filled  symbols}) the  source frame  peak
energy  $E_{peak}(1+z)$   versus  the  collimation   corrected  energy
$E_{\gamma}$ for  the 24 GRBs with  redshift and  $\theta$.  We
find that these two parameters are highly correlated with a Spearman's
rank correlation  coefficient $r_{s}$=0.88 and  a null hypothesis
probability of  being uncorrelated of  2.7$\times$10$^{-8}$.  If
we exclude  from the correlation all  the bursts (8)  for which either
the computed $E_{peak}$ and/or  $E_{\gamma}$ are upper/lower limits we
derive  a   correlation  coefficient  of   $r_{s}$=0.94  (with  a
probability of  1.4$\times$10$^{-7}$) and the  best fit powerlaw
model,  obtained accounting  for errors  on both  coordinates (routine
fitexy of Press et al.  1999), is ({\it solid line} in Fig.~1)
\begin{equation} 
E_{peak}=267.0\,\left(E_{\gamma}\over{4.3\times 10^{50}\,  
{\rm erg}}\right)^{0.706\pm0.047} \, \, \, {\rm keV} \label{bella} 
\end{equation} 
Note that fitting $E_{\gamma}$ versus $E_{peak}$ with this method 
gives $E_{\gamma}\propto  E_{peak}^{1.416\pm0.09} $, which  is exactly 
equivalent to Eq. \ref{bella}. Fig.~1 also shows ({\it open symbols}), 
for  all the  30  GRBs present  in  Tab.~1, the  peak  energy and  the 
isotropic   equivalent  energy  $E_{\gamma,iso}$.    This  correlation 
presents a larger  spread with respect to previous  findings (Amati et 
al.   2002, Lamb  et al.  2003) due  to the  larger number  of objects 
included   in  the  sample.    The  correlation   coefficient  results 
$r_{s}$=0.803 (with a probability of 7.6$\times$10$^{-7}$) and 
excluding the upper/lower  limits on $E_{peak}$ (3 GRBs)  the best fit 
is  $E_{peak}=258 (E_{\gamma,iso}/1.2\times10^{53} erg)^{0.40\pm0.05}$ 
({\it   dashed  line}   in   Fig.~1).   Note   that,  especially   for 
$E_{peak}\sim$ 500--600  keV, there is  a relatively large  scatter of 
$E_{\gamma,iso}$ values  around the fitting  line.  For the  very same 
bursts    the    scatter   of    $E_\gamma$    around   the    fitting 
$E_{peak}-E_{\gamma}$  line is much  reduced.  This  means that  it is 
possible  that bursts  with the  same $E_\gamma$  and  $E_{peak}$ have 
different jet opening angle.

Of all  the bursts  included in Tab.~1  only GRB~980425  represents an
outlier for the $E_{peak}$--$E_{\gamma,iso}$ correlation, and for this
burst  the angle  is  unknown.  Instead  XRF030723  and XRF020903  are
consistent with the extrapolation of the above correlation at very low
peak energies  (see also Lamb et  al.  2003). For  XRF030723 there are
indications  of a  break in  the lightcurve  between 30  and  50 hours
(Dullighan et  al. 2003),  while for  XRF020903 a jet  break time
around 3 days  (corresponding to $\sim25$ days), would  be required to
bring it  on the  $E_{peak}$--$E_{\gamma}$ correlation.  The  8.46 GHz
data  of  XRF020903 presented  in  Soderberg  et  al.  (2004)  show  a
relatively fast  time decay (decay  index $< -1.5$), after  the source
became transparent (i.e.  $\sim$25  days after the trigger).  However,
Soderberg et  al. (2004) claim that there  is no sign of  jet break at
4.86  GHz, where  the flux  increases up  to $\sim$30  days  after the
trigger. If this is the case, then  XRF020903 would be an outlier.

The  correlation  that  we  find  between  the  peak  energy  and  the
collimation corrected energy is also extremely narrow and we note that
except for one burst (GRB~000911) all the upper/lower limits are still
consistent  with  this  correlation.    In  Fig.~2(a)  we  report  the
distribution of the isotropic energy which, similarly to what reported
by  many authors  (Bloom et  al. 2001,  Frail et  al. 2001,  Berger et
al. 2003),  is wide spread over  almost 3 orders  of magnitude between
6.8$\times$10$^{51}$ erg  and 2.8$\times$10$^{54}$ erg.   Applying the
correction  for  the  collimation  angle, this  distribution  clusters
around  a  typical  value  of $\sim$10$^{51}$  erg  (Fig.~2(b)).   The
gaussian fit  to the logarithmic distribution  of $E_{\gamma}$ results
in a central  value $\langle \log(E_{\gamma})\rangle$=50.8 with a
standard deviation of 0.6. Introducing the correlation of $E_{\gamma}$
with $E_{peak}$ we can derive a  measure of the dispersion of the GRBs
around  the  correlation  $E_{peak}\propto  E_{\gamma}^{0.7}$  in  the
log--log  plane.   The  distribution  of  the  dispersion  measure  is
reported in Fig.~2(c) and its average value is $0.041\pm 0.015$ with a
maximum   logarithmic  dispersion   of  0.25.    We  note   that  this
distribution  represents  the   effective  tight  correlation  between
$E_{peak}$ and $E_{\gamma}$.

\subsection{Consistency  checks} 
 
From the  sample of  GRBs with measured  redshift we excluded  8 cases 
with uncertain  redshift and/or peak  energy. These GRBs  are reported 
(with the same meaning of the  columns) in Tab.~3.  In this section we 
discuss their consistency with  the above correlation under reasonable 
assumptions on their lacking parameters. 
 
GRB~970508 was  detected both by  $Beppo$SAX and BATSE.   According to 
the analysis of $Beppo$SAX data (Amati  et al 2002) its peak energy should be 
145$\pm$43 keV whereas the analysis  of the BATSE spectrum (Jimenez et 
al.   2001)  reports an  unconstrained  peak  energy  (due to  a  high 
powerlaw  spectral index  $\beta>$-2).  The  latter determines  only a 
lower  limit of $E_{peak}>$1503  keV (rest  frame).  A  more difficult 
situation was encountered for 5 GRBs (991208, 000210, 000301C, 000418, 
000926) with unpublished  spectral parameters.  Except  for GRB991208, 
the  other  events were  detected  by one  of  the  satellites of  the 
Interplanetary Network.   These instruments are  most likely sensitive 
to  GRBs  with  a spectrum  similar  to  that  of bursts  detected  by 
BATSE. For  this reasons, having only  a fluence, we  assumed a typical 
Band function with $\alpha$ and $\beta$ equal to the average values of 
these parameters for  the bursts presented in Tab.~1  (see also sec.2) 
and with a break energy free to vary in the range 100-500 keV which is 
a conservative  estimate of the  typical width of the  distribution of 
this spectral parameter for BATSE  bright bursts (Preece et al. 2000). 
With the above assumptions we iteratively calculated $E_{\gamma,iso}$, 
$E_{peak}$ and $E_{\gamma}$ for these bursts and report their interval 
of variation in Fig.~3.  In  Tab.~4 we report for reference their peak 
energy, isotropic  equivalent energy and  collimation corrected energy 
corresponding  to the assumption  of the  above average  spectrum with 
break  energy equal  to  the central  value  of its  allowed range  of 
variation.    
 
Finally there are two GRBs (980326 and 980329) with uncertain redshift 
measurements.  There  are some indications  (Bloom et al.   1999) that 
980326 lies  at $z<$1.5 (this limiting  value was assumed  by B03) and 
that 980329 might be between redshift 2 and 3.9.  For these two events 
we         calculated         the         region        of         the 
$E_{peak}$--$E_{\gamma,iso},E_{\gamma}$  plane were  they  lie varying 
their $z$ within these ranges.   They are represented by the connected 
polygons  in Fig.~3.   In Tab.~4  we report  for these  two  cases the 
parameters obtained assuming  the median values (i.e. 1.0  and 3.0) of 
the ranges where their redshift was allowed to vary. 
 
We can  see from Fig.~3 that  with the above  reasonable assumptions on 
the uncertainty of their spectral  parameters or distance also these 8 
GRBs  are   still  consistent  with  the   found  correlation  between 
$E_{peak}$ and $E_{\gamma}$.

\section{Discussion} 
 
The  analysis of  a sample  of GRBs  with measured  redshift  and peak 
energy revealed a tight correlation between their peak spectral energy 
and collimation corrected energy: 
\begin{equation} 
E_{peak}   \,   \simeq   480   \,   
\left({E_\gamma\over   10^{51}{\rm erg}}\right)^{0.7} \,\, {\rm keV} 
\label{23} 
\end{equation} 
The maximum (logarithmic) scatter  from this correlation in the sample 
of 24 GRBs with measured  jet opening angle is $\sim$0.25~dex while its 
distribution  (Fig.~2(c), insert)  has  a standard  deviation of  less than 
0.1~dex. 
 
The distribution of  the collimation--corrected energy (Fig.~2(b)) for 
our  GRB  sample  presents  a   wider  spread  compared  to  the  same 
distribution in F01 and B03. This is due to the different energy bands 
in which  the $E_\gamma$ has  been computed and  to the fact  that our 
sample  includes several  recent HETE-2  GRBs not  present in  the B03 
sample (nor obviously in the F01 one).  Nevertheless the central value 
of our distribution $\sim$  6$\times$10$^{50}$ erg is consistent with 
what reported  in F01.  The  tight correlation of Eq.~\ref{23}  can be 
found in both the F01 and the B03 samples, even though the peak energy 
range is smaller and therefore the statistical significance lower. 
 
We derived  the jet opening angle  (Eq.~1) within the  framework of the 
standard afterglow  theory (Sari et al.  1999).   The small scattering 
around the correlation between $E_{peak}$ and $E_{\gamma}$ can be seen 
as  an {\it  a-posteriori}  check  that the  assumptions  we made  are 
correct.  This indicates  that  the circum-burst  density, its  radial 
profile  and  the  efficiency  coefficient  $\eta_\gamma$  are  fairly 
standard. 
 
We have checked that no burst among the 8 cases reported in Tab.~3 and
Tab.~4, for  which we  have only a  partial knowledge of  the relevant
parameters,  contradicts the  correlation (Fig.~3).    One X--ray
flash  with  known redshift  (020903)  could  be  consistent with  our
correlation    assuming    a    jet   opening    angle    $\theta\sim$
25$^{\circ}$. The  existing radio data are controversial  and this XRF
could be an  outlier.  The other XRF in our  sample (030723) lies on
our   correlation  if   its  redshift   is  not   much   smaller  than
unity.  Concerning the two  GRBs associated  with a  SN, we  have that
GRB~980425/SN1998bw  (Galama et al.   1998) has  a very  low isotropic
energy and no  jet break and it is standing alone  with respect to any
other GRB in  the $E_{peak}$--$E_{\gamma}$,$E_{\gamma,iso}$ plane.  On
the other  hand GRB~030329/SN2003dh  (Stanek et al.  2003) with  a jet
opening angle of $\sim$5$^{\circ}$ (Price et al.  2003a, Tiengo et al.
2003, Berger  et al.  2004) has  a collimation corrected  energy and a
peak  spectral  energy  which  place on  the  $E_{peak}$--$E_{\gamma}$
correlation.
 
The sample  of 29  GRBs (Tab.~1,~2) also  allowed to  re--consider the 
correlation between  the isotropic equivalent  energy $E_{\gamma,iso}$ 
and the peak  energy $E_{peak}$ found by Amati et  al. (2002; see also 
Lloyd-Ronning et al.  2000, Lamb et al.  2003; Sakamoto et al.  2004). 
Including  all GRBs  with measured  redshift (Tab.~1)  we confirm the  
existence of  such correlation.   However, we 
find  two differences.  On the  one hand  the spread  around  the best 
correlation  line is  larger  than previously  estimated with  smaller 
samples.  On the  other hand,  the slope  of the  correlation  seem to 
depend on whether  XRFs are included or not.  We find that considering 
only            GRBs,            the            relation            is 
$E_{peak}\propto{}E_{\gamma,iso}^{0.40\pm0.05}$,      flatter     than 
previously estimated,  shown with a  {\it dashed line} in  Fig.~1 (see 
also  Amati  2004, who  finds  a  slope  $0.45\pm0.06$).  If  instead, 
following Lamb et al.  (2003),  we include X-ray flashes, the original 
Amati  et al.   (2002) result  fits better  the data  ({\it dot-dashed 
line} in Fig.~1).   The paucity of GRB/XRFs with  peak energies in the 
tens of keV range does not  allow us to draw any firm conclusion (note 
that the two slopes are consistent at the 2$-\sigma$ level). 
 
The  different slopes of  the $E_{peak}$--$E_{\gamma,iso}$ (Amati 
et al.  2002) and the $E_{peak}$--$E_{\gamma}$ correlations imply that 
they will  intersect at some small  value of $E_\gamma=E_{\gamma,iso}$ 
(see Fig.1).  The precise value  of the intersection depends mainly on 
the  uncertainties in  the  $E_{peak}$--$E_{\gamma,iso}$ relation,  as 
illustrated in  Fig. 1 (dashed  and dot--dashed lines).  At  any rate, 
bursts  lying  in   this  region  of  the  $E_{peak}$--$E_{\gamma}\sim 
E_{\gamma,iso}$ plane would be  characterized, on average, by a nearly 
isotropic   emission,   small   peak   energies   (possibly   in   the 
UV/soft--X--ray range) and by $E_\gamma$ smaller than $10^{47}$ erg 
\footnote{Note  that GRB 980425 is  an outlier whether  or not its 
$\gamma$--ray emission was isotropic.}.

Recently,  it  has  also been  shown  (Liang  et  al. 2004)  that  the 
correlation  between $E_{peak}$ and  $E_{\gamma,iso}$ (found  from the 
time-averaged  spectrum)  holds  when  considering the  time  resolved 
spectral analysis of  GRBs.  Similarly, due to the  much lower scatter 
(a  factor $\sim$ 5.7  lower) around  the $E_{peak}$  and $E_{\gamma}$ 
correlation,  we expect  that considering  the time  resolved spectral 
peak   energy   $E_{peak}(t)$   and   collimation   corrected   energy 
$E_{\gamma}(t)$  a possibly  even less  scattered  correlation should 
hold within single  GRBs.  This, in turn, might  help in exploring the 
(still obscure) origin  of the prompt emission.  In  fact, although we 
do not  have an  interpretation of the  found $E_{peak}$--$E_{\gamma}$ 
correlation,  we  believe  it  indicates  that there  should  be  some 
connection between  the energy emitted  by the burst and  the emission 
process which determines its spectral properties. 
 
The existence of the above  correlation also allows to predict the jet 
break time, once  the spectrum and the redshift of  a burst are known. 
In the forthcoming  Swift era this might contribute  in optimizing the 
GRB  follow up  with planned  observations at  the expected  jet break 
time.   For  this  reason we  think  that  it  is important  that  the 
information about  the peak  energy be disseminated  promptly together 
with the other  fundamental GRB properties such as  the position, peak 
flux    etc.     Now   this    routinely    occurs    for   HETE    II 
bursts\footnote{http://space.mit.edu/HETE/Bursts/}, and  it is crucial 
that this will happen also for the future Swift bursts. 
 
Berger et al. (2004a) recently  discussed the possibility that the real 
standard  energy  in  cosmic  explosions  (including  GRBs,  XRFs  and 
hypernovae)  is the  total  released kinetic  energy  rather than  the 
fraction that  goes into  $\gamma$-ray photons.  This  energy includes 
corrections due to mildly  relativistic material (estimated from radio 
emission such  as in SN1998bw; Kulkarni  et al. 1998) and  to a longer 
time-scale  activity  of  the  engine (estimated  from  the  afterglow 
lightcurve, such as  in GRB~030329).   
While not in contrast  with this idea, our findings underline the 
fact that the $\gamma$-ray  energetics is still an important parameter 
for understanding the physics  of GRBs.  The found correlation implies 
that the peak energy of the spectrum, which is a quantity set by local 
micro--physical processes,  is determined by a global  property of the 
jet, namely its total $\gamma$--ray energy (or the jet opening angle).

Last  but not  least,  the $E_{peak}$--$E_{\gamma}$  relation that  we 
presented in  this work and  the low scatter  of the sample  of bursts 
with  known redshifts around  it strongly  reminds a  similar relation 
found for SN Ia between  their luminosity and the stretching factor of 
their  optical light  curve  (e.g. Phillips  1993,  Perlmutter et  al. 
1998) with  less luminous  supernovae showing  a  faster post--maximum 
light curve decay (Riess et  al.  1995).  The proper modelling of this 
effect (e.g.  Hamuy  et al.  1996, Perlmutter et  al.  1998) allows to 
better determine the SN luminosity and consequently reduce the scatter 
in the Hubble  diagram with respect to the case  that assumes that all 
SN Ia  are standard  candles. Similarly, our  result shows  that among 
GRBs the existence  of a correlation with small  scatter between their 
peak  spectral energy and  the collimation  corrected energy  might be 
used to determine the different luminosities of different bursts.  The 
small   scatter  around   the   rest  frame   $E_{peak}$--$E_{\gamma}$ 
correlation might  be used  to reduce the  scatter of  the GRB--Hubble 
diagram (Bloom et al.  2003,  Schaefer et al.  2003) and constrain the 
cosmological   parameters  (Ghirlanda   et   al.   (2004)   in 
preparation).   If  our  relation  will  be  confirmed  and  possibly 
extended (particularly to low  peak energies) with new bursts detected 
by Swift, we might be able to test the cosmological models farther out 
the   predicted  SNAP   SN   limit  of   $z$=1.7   (e.g  Aldering   et 
al. 2002). This is particularly  interesting since we might be able to 
explore  an  epoch  in  which  the  universe,  matter  dominated,  was 
decelerating. 
 
\section{Conclusions} 
 
We considered all the bursts  with measured redshift and spectral peak 
energy up to  now (June 2004).  We systematically derived their source 
frame bolometric isotropic equivalent  energy and, when available from 
the  afterglow  modeling  (24/38),  corrected this  energy  for  their 
collimation  angle.  We  found a  very tight  correlation  between the 
source frame $E_{peak}$  (the peak in the $\nu  F_\nu$ spectrum of the 
prompt  emission)  and   the  collimation  corrected  emitted  energy: 
 $E_{peak} \sim 480 \, (E_\gamma/10^{51}{\rm erg})^{0.7}$ keV. 
 
\begin{itemize} 
 
\item The small dispersion around this correlation is an indication of 
the robustness  of the afterglow theory  on which the  estimate of the 
jet opening angle is based (Eq.~1, Sari et al. 1999) 
 
\item We  found that, differently  from previous results, GRBs  do not 
cluster around  a unique value of their  collimation corrected energy, 
but  are spread  in  a  relatively large  range  ($\sim$2 orders  of 
magnitudes)  centered at  $E_{\gamma}\sim$6$\times$10$^{50}$  erg.  We 
believe  that the  smaller spread  found previously  around  a similar 
value $\sim$5$\times$10$^{50}$  erg (Frail et al.  2001,  Bloom et al. 
2003,  Panaitescu \&  Kumar 2001)  might be  the result  of  a limited 
sample of GRBs within a relatively small $E_{peak}$ range. 
 
\item  The  underlying  physical  motivation  of  the  correlation  is 
mysterious, but  it is likely that  it is the result  of the radiation 
process producing the  prompt emission which should be  related to the 
local energy content of the burst. 
 
\item There are relatively few bursts with $10<E_{peak}<100$ keV, just 
the range of  the BAT instrument on-board Swift,  which therefore will 
be crucial for confirming or discarding the correlation. 
 
\item  Finally, and more  importantly, the  correlation (if  the small 
scatter around it will be confirmed) makes GRBs exquisite cosmological 
tools, to  measure $\Omega_m$, $\Omega_{\Lambda}$ in  a redshift range 
not  accessible to  Supernovae Ia,  that is,  $z>1.7$, and  up  to any 
redshift.  Even if the optical information is unavailable for $z>2.5$, 
the necessary information about the  jet break time can be gathered by 
following the X--ray and/or  the IR afterglow.  Partial obscuration by 
dust, which can be an  issue for using Supernovae as standard candles, 
does not influence GRBs. 
 
\end{itemize}

\acknowledgments{We thank the referee for her/his comments which 
improved the paper and Annalisa Celotti for useful discussions.  DL 
thanks the Osservatorio Astronomico  di Brera for the kind hospitality 
during part of the preparation  of this work.  G. Ghirlanda thanks the 
ASI (/I/R/390/02) for financial support. We thank M. R. Panzera for  
technical support. }

\clearpage

\begin{deluxetable}{lllllllll} 
\tabletypesize{\footnotesize} 
\tablecaption{Sample of GRBs\label{tbl-1}} 
\tablewidth{0pt} 
\tablehead{ 
\colhead{GRB$^a$}  & \colhead{$z^b$} & \colhead{$\alpha$} & \colhead{$\beta$} & 
\colhead{$E_{peak}$ } & \colhead{T$_{90}$ } & 
\colhead{Fluence} & \colhead{Range } &\colhead{ref$^d$}\\ 
  &  & &  & \colhead{(keV)$^c$} &  \colhead{ (s)} &\colhead{ (erg/cm$^2$)} & \colhead{ (keV)} & 
} 
 
\startdata 
970228	S/B   &	0.695 (H)    &	-1.54 (0.08)   &-2.5 (0.4)    &	115 (38)       & 80	&1.1e-5 (0.1)  & 40-700   &   (1),(31) 	   \\ 
970828	R/A(B)&	0.957 (H)    &	-0.70 [0.08]   &-2.07 [0.37]  &	298 [59]       & 146.59 &9.6e-5 [0.9]  & 20-2000  &   (2),(32) 	   \\ 
971214 	S/B   &	3.42 (H)     &	-0.76 (0.1)    &-2.7 (1.1)    &	155 (30)       & 35	&8.8e-6 (0.9)  & 40-700   &   (3),(31) 	   \\ 
980425	B/S   & 0.0085 (H)   &  -1.266 [0.13]  &  ...	      & 118 [24]       & 37.41  &3.8e-6 [0.4]  & 20-2000  &    (4),(32)	   \\ 
980613	S     &	1.096 (H)    &	-1.43 (0.24)   &-2.7 (0.6)    &	93 (43)        & 20     &1.0e-6 (0.2)  & 40-700   &   (5),(31)     \\ 
980703	R/A(B)& 0.966 (O)    &	-1.31 [0.14]   &-2.39 [0.26]  &	255  [51]      & 102.37 &2.3e-5 [0.2]  & 20-2000  &   (6),(32)     \\ 
990123	S/B   &	1.6 (O)      & 	-0.89 (0.08)   &-2.45 (0.97)  &	781 (62)       & 100    &3.0e-4 (0.4)  & 40-700   &   (7),(31)     \\ 
990506	B     &	1.3066	     &	-1.37 [0.15]   &-2.15 [0.38]  &	283 [57]       & 220.38 &1.9e-4 [0.2]  & 20-2000  &   (8),(32) 	   \\ 
990510	S/B   &	1.619 (O)    &	-1.23 (0.05)   &-2.7 (0.4)    &	163 (16)       & 75     &1.9e-5 (0.2)  & 40-700   &   (9),(31)     \\ 
990705	S     &	0.843 (XP,H) &	-1.05 (0.21)   &-2.2 (0.1)    &	189 (15)       & 42     &7.5e-5 (0.8)  & 40-700   &   (10),(31)	   \\ 
990712    S   &	0.43 (O,H)   &	-1.88 (0.07)   &-2.48 (0.56)  &	65 (11)        & 20     &6.5e-6 (0.3)  & 40-700   &   (11),(31)	   \\ 
991216	 B    &	1.02 (O)     &	-1.234 [0.13]  &-2.18 [0.39]  &	318 [64]       & 24.9   &1.9e-4 [0.2]  & 20-2000  &   (12),(32)    \\ 
000131	B     &	4.5 (H)      &	-0.688 [0.08]  &-2.07 [0.37]  &	130 [26]       & 110.1  &4.2e-5 [0.4]  & 20-2000  &   (13),(32)    \\    
000214	S     &	0.42(XP)      &	-1.62 (0.13)   &-2.1          &	$>$82          & 10     &1.4e-5 (0.04) & 40-700	  &   (14),(31)	   \\ 
000911	I     &	1.058 (H)    &	-1.11 [0.12]   &-2.32 [0.41]  &	579 [116]      & 500    &2.2e-4  [0.2] & 15-8000  &   (15),(33)    \\ 
010222    S   &	1.473 (O)    &	-1.35 (0.19)   &-1.64 (0.02)  &	$>$358	       & 130    &9.3e-5 (0.3)  & 40-700   &   (16),(31)    \\    
010921	H     &	0.45 (H)     &	-1.49 [0.16]   &-2.3          &	106 [21]       & 24.6   &1.0e-5 [0.1]  & 30-700   &   (17),(34,35) \\ 
011121	S     &	0.36 (O)     &	 ...           &... ($>$-2)   &	$>$700         & 75.    &1.0e-4 [0.1]  & 40-700   &   (18),(36)    \\ 
011211 S      & 2.14 (O)     &  -0.84 (0.09)   &...           & 59 (7)         & ...    &...	       & ...	  &   (19),(31)	   \\ 
020124	H     &	3.2 (O)      &	-1. [0.11]     &-2.3 [0.41]   &	110 [22]       &78.6    &6.8e-6  [0.7] & 30-400   &   (20),(35)    \\ 
020405	I/S   &	0.69 (O)     &	-0.0 (0.25)    &-1.87 (0.23)  &	364 [73]       &60      &7.4e-5  [0.7] & 15-2000  &   (21),(37)	   \\    
020813	H     &	1.25 (H)     &	-1.05 [0.11]   &-2.3          &	211 [42]       &90.     &1.0e-4	[0.1]  & 30-400   &   (22),(34,35) \\ 
020903X H     & 0.25 (H)     &  ...	       &...           & 5.52	       &...	&...	       & ...	  &    (23),(34)   \\ 
021211 	H     &	1.01 (O)     &	-0.85 [0.09]   &-2.37 [0.42]  &	47 [9]         &2.41    &2.2e-6 [0.2]  & 30-400   &   (24),(38)	   \\ 
030226	H     &	1.98 (O)     &	-0.95 [0.10]   &-2.3          &	108 [22]       &76.8    &6.4e-6  [0.6] & 30-400   &   (25),(34,35) \\ 
030328	H     &	1.52 (O)     &	-1.0 [0.11]    &-2.3	      &	110 [22]       &140     &2.6e-5  [0.2] & 30-400   &   (26),(34,35) \\ 
030329	H     &	0.1685 (O)   &	-1.26 (0.02)   &-2.28 (0.05)  &	68 (2)         &23.     &1.1e-4  [0.1] & 30-400   &   (27),(39)	   \\    
030429	H     &	2.66 (O)     &	...	       &...	      &	35 [7]         &10.3    &3.8e-7  [0.3] & 30-400   &   (28),(40)	   \\ 
030723X H     & $<$ 2.1      &  ...   	       &...	      &  4.8	       &...	&....	       & ...	  &   (29),(41)	   \\   	       			 
\enddata 
\tablenotetext{'}{In round parenthesis are reported errors actually measured; when these are unavailable we assume the errors reported in square parenthesis. } 
\tablenotetext{a}{Instrument(s) that were triggered by the GRB: S=$Beppo$SAX; B=BATSE; (B)=also BATSE, R=RXTE, A=ASM, I=IPN (Ulysses, NEAR, Konus),  
H=Hete-II. X--Ray Flashes (XRF) are indicated with the symbol X. } 
\tablenotetext{b}{Redshift determined from: H=Host Galaxy spectrum; XP=X--Ray Photometric data; O=Optical Transient.} 
\tablenotetext{c}{Observed peak energy.} 
\tablenotetext{d}{References given in order for: redshift, spectral informations.} 
 
\tablerefs{ 
  (1) Djorgovsky  et  al.  1998;  (2)  Djorgovsky  et  al.  2001;  (3) 
  Kulkarni et al. 1998; (4) Tinney  et al.  1998; (5) Djorgovsky et 
  al.  1998a;  (6)Djorgovsky et al.   1998b; (7) Hjorth   1999; (8) 
  Bloom et al.   2001; (9) Vreeswijk  et 
  al.  1999; (10) Amati  et  al.  2000; (11) Galama  et al. 1999; 
  (12) Vreeswijk  et  al.   1999a;  (13)  Andersen et  al.  2000;  (14) 
  Antonelli et al.  2000; (15) Price et al.  2002; (16) Stanek et 
  al. 2001; (17)  Djorgovski et al.  2001a; (18)  Infante et al.  2001; 
  (19) Amati   2004;  (20) Hjorth  et  al.  2003;  (21) Masetti  et 
  al. 2002; (22) Price et al.  2002a; (23) Soderberg et al.  2004; (24) 
  Vreeswijk et  al.  2003; (25) Greiner  et al. 2002; (26)  Rol et al. 
  2003; (27)  Greiner et al.  2003;  (28) Weidinger et  al. 2003; (29) 
  Fynbo et  al.  2004; (30) Berger  et al. 2004b;  (31) Amati 
  et al.   2002; (32)Jimenez et al.  2001;  (33) Price et al.  2002; 
  (34) Barraud et al.  2003; (35) Atteia et al.  2003; (36) Piro et al. 
  2004;  (37)  Price  et  al.  2003;  (38) Crew  et  al.   2003;  (39) 
  Vanderspek  et  al.   2004; (40)  http://space.mit.edu/HETE/Bursts/; 
  (41) Lamb et al. 2003 } 
 
\end{deluxetable} 
 
\clearpage 
 
\begin{deluxetable}{llllllll} 
\tabletypesize{\small} 
\tablecaption{Sample of GRBs\label{tbl-2}} 
\tablewidth{0pt} 
\tablehead{ 
\colhead{GRB}  &  
\colhead{$t_{jet}$ } & \colhead{$n$} & \colhead{ref$^a$} &\colhead{$\theta$} & \colhead{$E_{\gamma,iso}$} & 
\colhead{$E_{\gamma}$} & \colhead{$E_{peak}(1+z)$ }\\ 
 & \colhead{(days)} & \colhead{(cm$^{-3}$)} &  &\colhead{(deg)} & \colhead{(erg)} & \colhead{(erg)} & \colhead{(keV)} 
} 
 
\startdata 
970228  &	...         &	...         &	...,...	   &   ...	   &  1.60e52 (0.12)  & ...	         & 195   (64)	      \\ 
970828	&	2.2 (0.4)   &  	3.0 [1-10]  &	(1),...    &  5.9 [0.8]    &  2.96e53 [0.35]  & 1.6e51 [0.5]     & 583   [116]	      \\ 
971214 	&	$>$2.5 [0.4]&	3.0  [1-10] &	(1),...    &  $>$4.77	   &  2.11e53 (0.24)  & $>$7.29e50       & 685   (133)        \\ 
980425	&	...	    &    ...        &	 ...,...   &  ...	   &  1.6e48  [0.2]   & ....	         & 119   [24]	      \\ 
980613	&	$>$3.1	    &	3.0  [1-10] &	(1),...    &  $>$10.69	   &  6.9e51  (0.95)  & $>$1.0e50        & 194   (89)	      \\ 
980703	&	3.4 (0.5)   &	28 (10)     &	(1),(1)    &  11 [0.8]     &  6.9e52  [0.82]  & 1.27e51 [0.24]   & 502   [100]	      \\ 
990123	&	2.04 (0.46) &	3.0 [1-10]  &	(1),...    &  3.98  [0.57] &  2.39e54 (0.28)  & 5.76e51 [1.8]    & 2030  (161)      \\ 
990506	&	...         &	...	    &	...,...	   &   ...	   &  9.49e53 [1.13]  & ...	         & 653   [130]	      \\ 
990510	&	1.6 (0.2)  &	0.29 (0.1)  &	(15),(1)    &  3.74  [0.24] &  1.78e53 [0.19]  & 3.79e50  [0.63]  & 423   (42)	      \\ 
990705	&	1.0 (0.2)   &	3.0  [1-10] &	(1),...    &  4.78 [0.66]  &  1.82e53 (0.23)  & 6.33e50 [1.92]   & 348   (28)	      \\ 
990712  &	1.6 [0.2]   &	3.0  [1-10] &	(2),...    &  9.47  [1.2]  &  6.72e51  (1.29) & 9.16e49  [2.9]   & 93    (15)	      \\ 
991216	&	1.2 (0.4)   &	4.7 (2.3)   &	(1),(1)    &  4.4  [0.6]   &  6.75e53 [0.81]  & 2.0e51 [0.6]     & 641   [128]	      \\ 
000131	&	$<$3.5	    &	3.0 [1-10]  &	(1),...    &  $<$3.8	   &  1.84e54 [0.22]  & $<$4.04e51       & 714   [142]	      \\    
000214	&	...         &	...         &	...,...	   &   ...	   &  8.0e51  (0.26)  & ...	         & $>$117	      \\ 
000911	&	$<$1.5	    &	3.0  [1-10] &	(3),...    &  $<$4.4	   &  8.8e53  [1.05]  & $<$2.6e51        & 1190  [238]        \\ 
010222  &	0.93 (0.1)  &	1.7 [0.18]  &	(1),(1)    &  3.03  [0.14] &  1.33e54 (0.15)  & 1.85e51  [0.27]  & $>$886	      \\    
010921	&	$<$33.      &	3.0  [1-10] &	(1),...    &  $<$28.26	   &  9.0e51  [1.0]   & $<$1.e51	 & 153   [31]         \\ 
011121	&	$>$7	    &	3.0  [1-10] &	(1),...    &  $>$13.21	   &  4.55e52 [0.54]  & $>$1.2e51        & $>$952	      \\ 
011211  &       1.5 (0.15$^b$) & 3.0 [1-10] &   (4),...    &  5.2  [0.63]  &  6.3e52  (0.7)   & 2.5d50  [0.69]   & 186   (24)	      \\ 
020124  &	3. [0.4]    &	3.0  [1-10] &	(5),...    &  5.0 [0.3]    &  3.02e53 [0.36]  & 1.1e51 [0.32]    & 503   [100]	      \\ 
020405	&	1.67 (0.52) &	3.0  [1-10] &	(1),...    &  6.4 [1.05]   &  1.1e53  [0.13]  & 6.8e50 [2.39]    & 612   [122]	      \\    
020813	&	0.43 (0.06) &	3.0 [1-10]  &	(1),...    &  2.7 [0.13]   &  8.e53   [0.96]  & 8.8e50 [2.5]     & 474   [95]	      \\ 
020903X & 	...	    & 	...	    &    ...,...   &  ... 	   &  2.3e49	      & ...		 & $<$6.3	      \\ 
021211  &	$>$1	    &	3.0  [1-10] &	(6),...    &  $>$6.57	   &  1.1e52  [0.13]  & $>$7.2e49        & 94    [19]	      \\ 
030226	&	0.84 (0.1)  &	3.0 [1-10]  &	(12),(7)   &  3.94 [0.49]  &  1.2e53  [0.13]  & 2.83e50 [0.77]   & 322   [64]	      \\ 
030328	&	0.8 [0.1]   &	3.0  [1-10] &	(8),...    &  3.7  [0.46]  &  2.8e53  [0.33]  & 5.9e50 [1.6]     & 277   [55]	      \\ 
030329  &	0.5 (0.1)   &	1 [0.11]    &	(9),(10)   &  5.1  [0.4]   &  1.8e52  [0.21]  & 7.1e49 [1.4]     & 79    (3)	      \\    
030429	&	1.77 (1.0)  &	3.0 [1-10]  &	(13),...   &  5.96 [1.43]  &  2.19e52 [0.26]  & 1.2e50 [0.59]    & 128   [26]	      \\ 
030723X & 	1.67 [0.3]  &   3.0 [1-10]  &   (14),...   &  $>$10.6      &  $<$2.16e50      & $<$4.0e48  & $<$15   \\ 
\enddata  \tablenotetext{a}{References given in  order for:  jet break 
time,  external medium  density} 
\tablenotetext{b}{We have assumed that the error is 10\% instead of 
the 1.3\% error quoted in Jakobsson et al.  2003.} 
\tablerefs{  (1) Bloom  et  al.  2003 
(references therein);  (2) Bjornsson  et al.  2001;  (3) Price  et al. 
2002; (4) Jakobsson et al.  2003;  (5) Torii et al. 2002, Gorosabel et 
al.   2002, Bloom  et al.   2002; (6)  Della Valle  et al.   2003; (7) 
Pandey  et al.  2004;  (8) Andersen  et al.  2003; Perterson  \& Price 
2003,  Burenin et al.   2003; (9)Berger  et al.  2004; (10)  Tiengo et 
al.  2003; (11)  Bersier  et al.  2004.;  (12)  Klose  et al. 2004; 
(13) Jakobsson et al. 2004; (14) Dulligan et al. 2003 ; (15) Israel et al. 1999}  
\end{deluxetable} 
 
\newpage 
\clearpage 
\begin{figure} 
\includegraphics{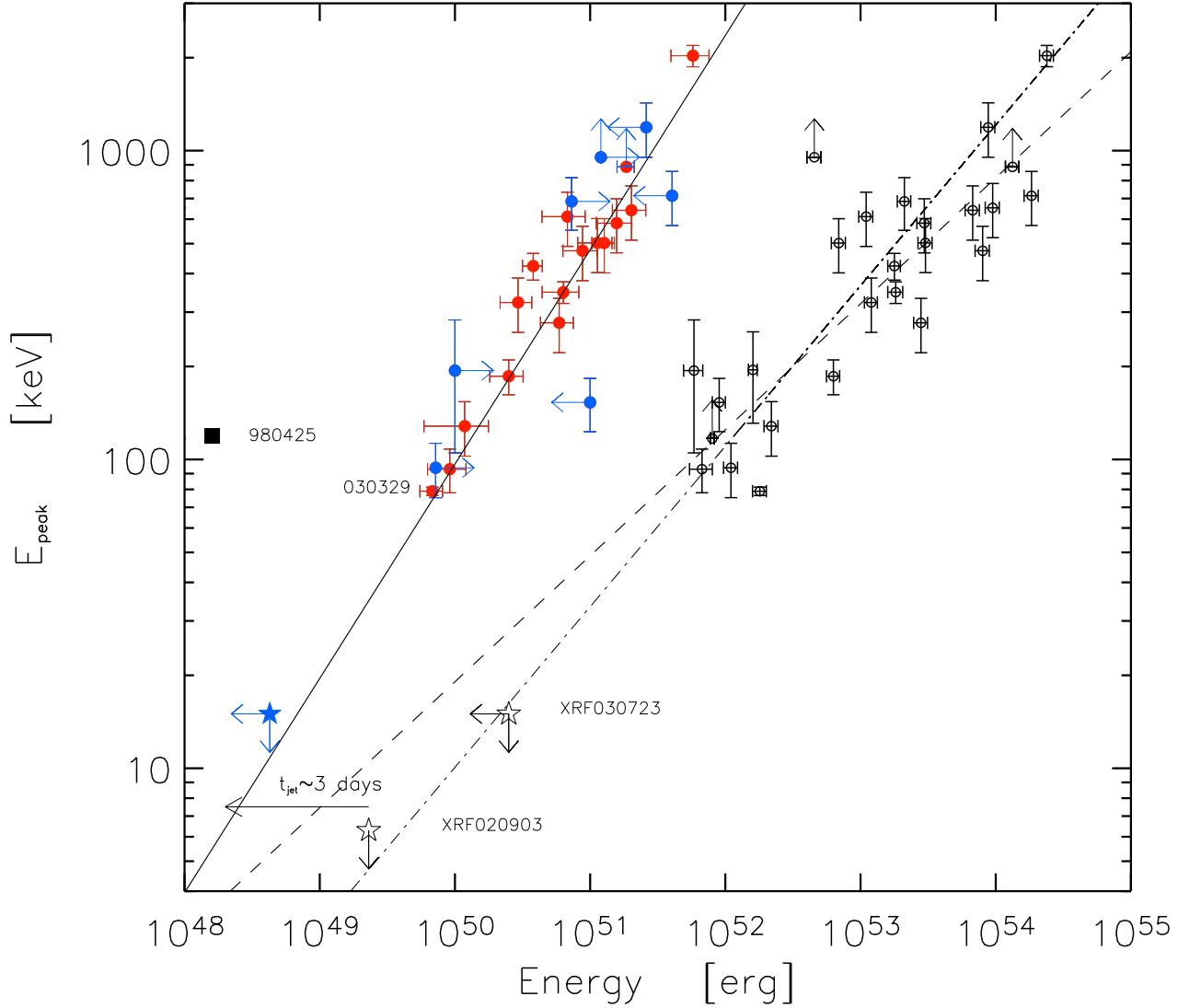}     
\figcaption[]{Rest   frame    peak   energy 
$E_{peak}=E_{peak}^{obs}(1+z)$ versus bolometric energy for the sample 
of  GRBs  with measured  redshift  reported  in  Tab.~1.  {\it  Filled 
circles}: isotropic energy corrected  for the collimation angle by the 
factor $(1-{\rm cos}\theta)$, for the  events for which a jet break in 
the light  curve was observed (see Tab.~2).   Grey symbols corresponds 
to lower/upper  limits.  The {\it Solid  line} represents the best  fit to 
the  correlation, i.e.   $E_{peak} \sim  480  \, (E_\gamma/10^{51}{\rm 
erg})^{0.7}$  keV.  {\it  Open circles}:  isotropic  equivalent energy 
$E_{\gamma,iso}$ for the GRBs reported in Tab.~2. The {\it Dashed line} is 
the  best fit  to  these points  and the  {\it dash--dotted  line} is  the 
correlation reported by Amati et al. (2002).
\label{fig1}} 
\end{figure}

\clearpage 
\newpage 
\begin{figure} 
\resizebox{!}{\vsize}{\includegraphics{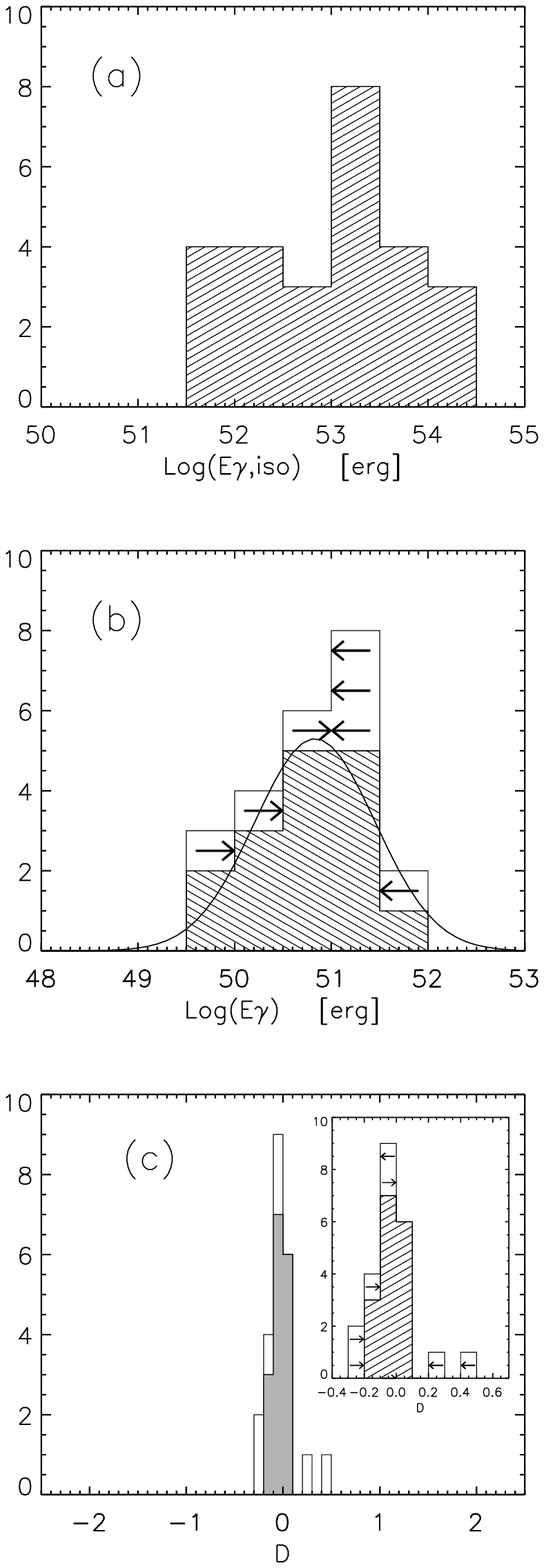}} 
\figcaption[]{Top: distribution of the rest frame isotropic equivalent 
energy $E_{\gamma,iso}$ for  the sample of GRBs presented  in Tab.~1. 
Middle:  Distribution  of   the  collimation--angle  corrected  energy 
$E_{\gamma}$. Bottom:  distribution of the  distance of each  GRB from 
the   correlation  between   $E_{peak}$  and   $E_{\gamma}$.   Insert: 
distribution of the distance from the correlation with upper and lower 
limits indicated.\label{fig2}} 
\end{figure} 
\clearpage

\begin{deluxetable}{lllllllll} 
\tabletypesize{\small} 
\tablecaption{Uncertain cases\label{tbl-3}} 
\tablewidth{0pt} 
\tablehead{ 
\colhead{GRB$^a$}  & \colhead{$z$} & \colhead{$\alpha$} & \colhead{$\beta$} & 
\colhead{$E_{peak}$ } & \colhead{T$_{90}$ } & 
\colhead{Fluence} & \colhead{Range } &\colhead{ref$^c$}\\ 
  &  & &  & \colhead{(keV)$^d$} &  \colhead{ (s)} &\colhead{ (erg/cm$^2$)} & \colhead{ (keV)} & 
} 
\startdata 
970508	S/B   &	0.835        &	-1.71 (0.1)   & -2.2 (0.25)    &	79 (23)     & 20	& 1.8e-6 (0.3)  & 40-700   &   (1),(7)   	 \\ 
              &              &  -1.19         & -1.83          &        $>1800$     &           & 5.5e-6 [0.5]  & 20-2000  &   ...,(8)           \\ 
980326  S/B     & ...          &  -1.23 (0.21)  & -2.48 (0.31)   &        33.8 (17.1) & 9         & 0.75e-6 (0.15)& 40-700   &   ...,(7)           \\ 
980329  S/B     & ...          &  -0.64 (0.14)  & -2.2  (0.8)    &       233.7 (37.5) & 25        & 6.5e-5  (5)   & 40-700   &   ...,(7)           \\ 
991208  I     & 0.7          &  ...           & ...            &       ...          & 60        & 1.e-4         & 25-1.e4  &   (2),(9)           \\  
000210  S     & 0.846        &  ...           & ...            &       ...          & 20        & 6.1e-5        & 2-700    &   (3),(10)           \\ 
000301C I     & 2.033        &  ...           & ...            &       ...          & 50        & 4.1e-6        & 25-1.e4  &   (4),(11)            \\ 
000418  I     & 1.118        &  ...           & ...            &       ...          & 30        & 2.e-5         & 15-1.e4  &   (5),(12)            \\ 
000926  I     & 2.036        &  ...           & ...            &       ...          & 25        & 2.2e-5        & 25-1.e4  &   (6),(13)           \\ 
\enddata 
\tablenotetext{a}{Instrument(s) that were triggered by the GRB: S=$Beppo$SAX; B=BATSE; (B)=also BATSE, R=RXTE, A=ASM, I=IPN (Ulysses, NEAR, Konus). } 
\tablenotetext{c}{References given in order for: redshift, spectral informations} 
\tablenotetext{d}{Observed Peak energy} 
\tablerefs{  
(1) Metzger  1997; (2) Dodonov    et al. 1999; (3) Piro et 
al. 2002; 4) Castro  2000;  (5) Bloom  et al.  2000; (6) 
Castro  2000a; (7) Amati et  al. 2002; (8) Jimenez et al. 2001; (9) 
Piran et  al. 2000;  (10) Stornelli et  al.  2000; (11)  Garnavich et 
al. 2000; (12) Hurley et al. 2000; (13) Hurley et al. 2000a. 
} 
\end{deluxetable} 

\begin{deluxetable}{llllllll} 
\tabletypesize{\small} 
\tablecaption{Uncertain cases\label{tbl-4}} 
\tablewidth{0pt} 
\tablehead{ 
\colhead{GRB$^a$}  &  
\colhead{$t_{jet}$ } & \colhead{$n$} & \colhead{ref$^e$} &\colhead{$\theta$} & \colhead{$E_{\gamma,iso}$} & 
\colhead{$E_{\gamma}$} & \colhead{$E_{peak}(1+z)$ }\\ 
 & \colhead{(days)} & \colhead{(cm$^{-3}$)} &  &\colhead{(deg)} & \colhead{(erg)} & \colhead{(erg)} & \colhead{(keV)} 
} 
\startdata 
970508	&	25 (5)  &  	3.0 [1-10]  &	(1),...    &  24.0 [3.3] &  7.1e51 (0.15)  & 6.1e50 [1.6]   & 145 [43]	      \\ 
	&	25 (5)  &  	3.0 [1-10]  &	(1),...    &  22.1 [3.05] &  1.4e52 [0.17]  & 1.02e51 [0.3]    & $>$1503 	      \\ 
980326	&	$<$0.4  &  	3.0 [1-10]  &	(1),...    &  $<$5.08     &  5.6e51 (1.1)   & $<$2.2e49       & 71 [36]	      \\ 
980329	&	$<$1    &  	20 [10]     &	(1),(1)    &  $<$3.33     &  2.1e54  (0.2)  & $<$3.5e51       & 935 (150)	      \\ 
991208	&	$<$2.1  &  	3.0 [1-10]  &	(1),...    &  $<$8.64     &  1.1e53         & $<$1.2e51       & ...	      \\ 
000210	&	$>$1.7  &	3.0 [1-10]  &	(1),...    &  $>$5.76     &  2e53           & $>$1e51         & ...	      \\ 
000301C	&	7.3(0.5)&  	27 (5)      &	(1),(1)    &  13.14       &  4.37e52        & 1.14e51         & ...	      \\ 
000418  &       25 (5)  &       27          &   (1),(1)    &  22.3        &  7.51e52        & 5.6e51          & ...           \\ 
000926  &       1.8(0.1)&       27 (3)      &   (1),(1)    &  6.19        &  2.7e53         & 1.6e51          & ...           \\ 
\enddata 
\tablenotetext{a}{References given in order for: jet break time and external medium density.} 
\tablerefs{ (1) Bloom et al. 2003. } 
 
\end{deluxetable} 
 
\clearpage 
 
\begin{figure} 
\includegraphics{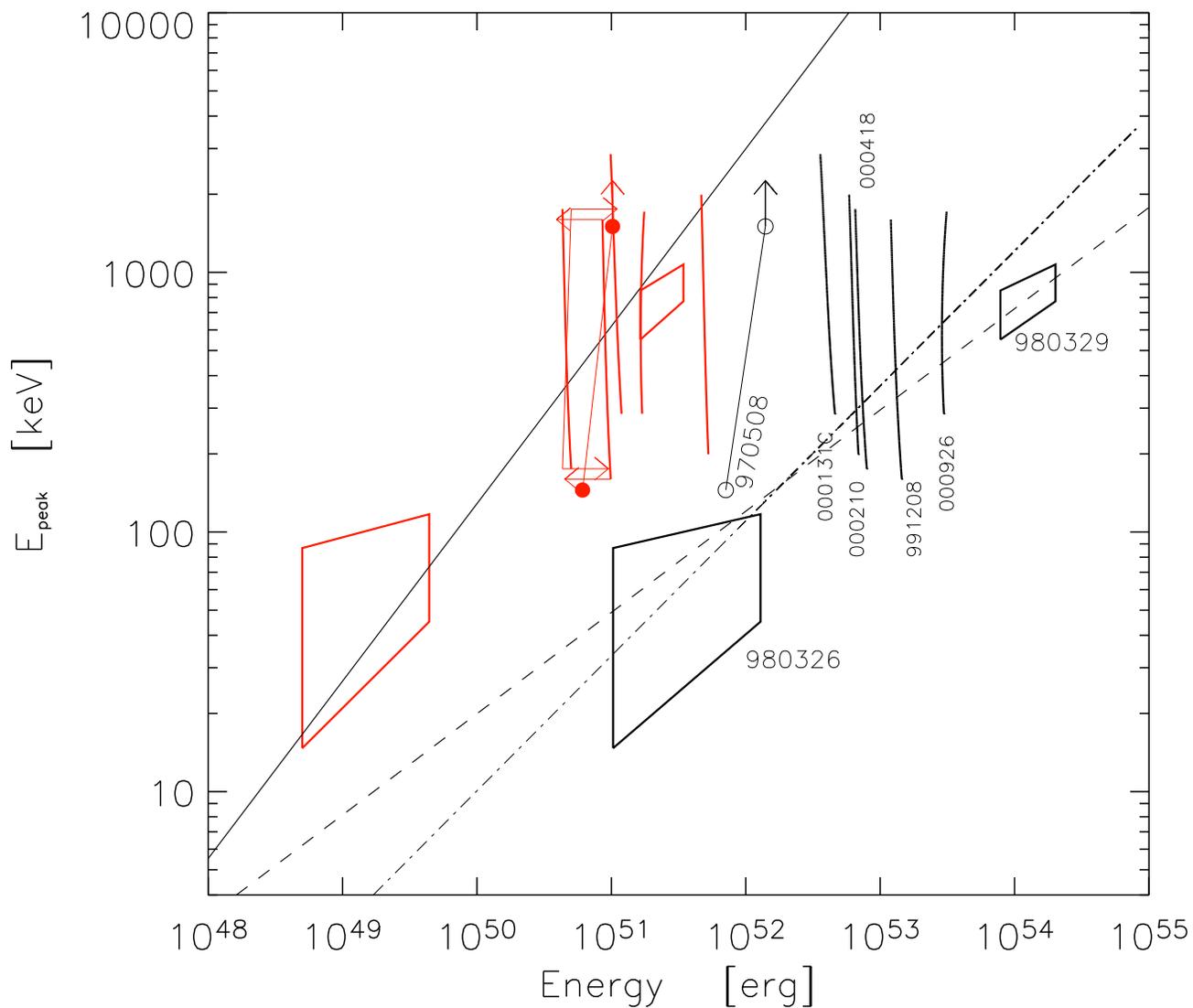}     
\figcaption[]{Rest   frame    peak   energy 
$E_{peak}=E_{peak}^{obs}(1+z)$ versus bolometric energy for the sample 
of GRBs reported  in Tab.~3,4 and excluded from  the larger sample due 
to (i) uncertain redshift (980329, 980326), (ii) uncertain peak energy 
(970508) or (iii)  no published spectral  information (991208, 000212, 
000131C, 000418, 000926). \label{fig3}} 
\end{figure}

\end{document}